# ROBUST PRECODER FOR MULTIUSER MISO DOWNLINK WITH SINR CONSTRAINTS

*P. Ubaidulla and A. Chockalingam*

Department of ECE, Indian Institute of Science, Bangalore 560012, INDIA

**ABSTRACT**

In this paper, we consider linear precoding with SINR constraints for the downlink of a multiuser MISO (multiple-input single-output) communication system in the presence of imperfect channel state information (CSI). The base station is equipped with multiple transmit antennas and each user terminal is equipped with a single receive antenna. We propose a robust design of linear precoder which transmits minimum power to provide the required SINR at the user terminals when the true channel state lies in a region of a given size around the channel state available at the transmitter. We show that this design problem can be formulated as a Second Order Cone Program (SOCP) which can be solved efficiently. We compare the performance of the proposed design with some of the robust designs reported in the literature. Simulation results show that the proposed robust design provides better performance with reduced complexity.

## 1. INTRODUCTION

There has been considerable interest in multiuser multiple-input multiple-output (MIMO) wireless communications in view of their potential for transmit diversity and increased channel capacity [1],[2]. Since it is difficult to provide mobile user terminals with large number of antennas due to space constraints, multiuser multiple-input single-output (MISO) wireless communications on the downlink, where the base station is equipped with multiple transmit antennas and each user terminal is equipped with a single receive antenna is of significant practical interest. In such multiuser MISO systems, multiuser interference at the receiver is a crucial issue. One way to deal with this interference issue is to use multiuser detection [3] at the receivers, which increases the receiver complexity. As an alternate way, transmit side processing in the form of precoding is being studied widely [2],[4]. Several linear precoders such as transmit zero-forcing (ZF) and minimum mean square error (MMSE) filters, and non-linear precoders including Tomlinson-Harashima precoder (THP) have been proposed and widely investigated in the literature [5],[6]. Both linear and nonlinear precoding strategies which meet SINR constraints at individual users have also been investigated [7],[8]. Non-linear precoding strategies, though more complex than the linear strategies, result in improved performance compared to linear pre-processing. Transmit side precoding techniques, linear or non-linear, can render the receiver side processing at the user terminal simpler. However, transmit side precoding techniques require channel state information (CSI) at the transmitter.

Several studies on transmit precoding assume perfect knowledge of CSI at the transmitter. However, in practice, CSI at the transmitter suffers from inaccuracies caused by errors in channel estimation and/or limited, delayed or erroneous feedback. The performance of precoding schemes is sensitive to such inaccuracies [9]. Several papers in the literature have proposed precoder designs, both linear and non-linear, which are robust in the presence of channel estimation errors [10],[11]. Linear robust precoding for MISO downlink with SINR constraints with imperfect CSI at the transmitter is considered in [12], where the robust design is formulated as Semi Definite Programs (SDP) of different performance and complexity. In [13], Payaro *et al.,* consider robust power allocation for fixed beamformers with Mean Square Error (MSE) constraints and formulate the problem as convex optimization problem. Robust power control for fixed beamformers with SINR constraints is considered in [14].

In this paper, we consider robust linear precoding under SINR constraints for the downlink of a multiuser MISO wireless communication system in the presence of imperfect CSI at the transmitter. The CSI at the transmitter is assumed to be perturbed by estimation or quantization error. The objective of the robust design considered here is to minimize the total transmit power, while ensuring the required SINR at each user. The robustness of the design consists in ensuring the target SINR for all errors in the CSI belonging to a given uncertainty region. We show that the design of the robust precoder with SINR constraints can be formulated as a convex optimization program. Specifically, we show that it is possible to formulate this problem as a SOCP which can be solved very efficiently. This is in contrast to the formulations in [12] which are computationally demanding. Our simulation results illustrate the improvement in performance and complexity compared to other robust precoders in the literature.

The rest of the paper is organized as follows. In section 2, we present the system model. The proposed robust precoder design is presented in section 3. Performance results and comparisons are presented in section 4. Conclusions are presented in section 5.

## 2. SYSTEM MODEL

We consider a multiuser MISO system, where a base station (BS) communicates with $N_u$ users on the downlink. The BS employs $N_t$ transmit antennas and each user is equipped with one receive antenna. Let $\mathbf{u}$ denote[1] the $N_u \times 1$ data symbol vector, where $u_i$, $i = 1, 2, \cdots, N_u$, denotes the complex valued data symbol meant for user $i$. The linear precoding matrix $\mathbf{B} \in \mathbb{C}^{N_t \times N_u}$ acts on this vector $\mathbf{u}$. The output of the precoding operation is denoted by the $N_t \times 1$ vector $\mathbf{x}$, where $x_j$, $j = 1, 2, \cdots, N_t$, denotes the complex-valued symbol transmitted on the $j$th transmit antenna. The received signal at user $i$, denoted by $y_i$, can be written as

$$y_i = \mathbf{h}_i \mathbf{B} \mathbf{u}_i + n_i, \qquad (1)$$

where $\mathbf{h}_i$ is the row vector containing complex channel gains from the transmit antennas to the receive antenna of user $i$, and $n_i$ is an i.i.d complex Gaussian random variable with zero mean and variance of $\sigma_n^2$ representing the noise at the $i$th receiver. The channel gains are assumed to be independent zero mean complex Gaussian variable and $E\{\mathbf{h}_i^H \mathbf{h}_i\} = \mathbf{I}$.

## 3. ROBUST PRECODER DESIGN WITH IMPERFECT CSI AT THE TRANSMITTER

In this section, we consider the design of robust precoder which transmits minimum power in order to provide the required SINR at each user when the CSI at the transmitter is imperfect. The SINR at user terminal $k$ is given by

$$\text{SINR}_k = \frac{|\mathbf{h}_k \mathbf{b}_k|^2}{\sum_{j=1, j\neq k}^{N_u} |\mathbf{h}_k \mathbf{b}_j|^2 + \sigma_k^2}, \qquad (2)$$

where $\mathbf{h}_k$ is the $k$th row of the matrix $\mathbf{H}$ and $\mathbf{b}_j$ is the $j$th column of matrix $\mathbf{B}$. Assuming $E\{\mathbf{u}\mathbf{u}^H\} = \mathbf{I}$, total transmit power is given by

$$P_T = E\{\mathbf{x}^H \mathbf{x}\} = \text{Tr}(\mathbf{B}^H \mathbf{B}) \qquad (3)$$
$$= \|\mathbf{b}\|^2, \qquad (4)$$

where $\mathbf{b} = \text{vec}(\mathbf{B})$.

When the transmitter has the perfect knowledge of CSI, the problem of designing a precoder which transmits minimum power while ensuring the required SINR at each user can be posed as

$$\min_{\mathbf{B}} \quad \text{trace}(\mathbf{B}^H \mathbf{B})$$
$$\text{subject to} \quad \text{SINR}_k \geq \gamma_k, \quad 1 \leq k \leq N_u, \qquad (5)$$

---
[1] Vectors are denoted by boldface lowercase letters, and matrices are denoted by boldface uppercase letters. $[.]^T$, $[.]^H$, and $[.]^\dagger$ denote transpose, Hermitian, and pseudo-inverse operations, respectively. $[\mathbf{A}]_{ij}$ denotes the element on the $i$th row and $j$th column of the matrix $\mathbf{A}$. vec(.) operator stacks the columns of the input matrix into one column-vector.

where $\gamma_k$ is the SINR required at the $k$th user. The above problem can be solved in different ways [8], [15]. The problem as stated in (5) is not convex. But it is shown in [15] that this problem can be formulated as a SOCP. This convex formulation enables the use of efficient numerical algorithms to solve the precoder design problem. The SOCP formulation of (5) is given by [16]

$$\min_{\mathbf{B}} \quad \tau$$
$$\text{subject to} \quad \|\mathbf{b}\| - \tau \leq 0, \qquad (6)$$
$$\left\|[\mathbf{h}_k \mathbf{B} \ \sigma_k]\right\| - a_k \mathbf{h}_k \mathbf{b}_k \leq 0, \quad 1 \leq k \leq N_u,$$

where $a_k = \sqrt{\frac{1}{1+\gamma_k}}$. Here, we have assumed that the imaginary part of $\mathbf{h}_k \mathbf{b}_k$ is zero. This is possible because we can add arbitrary phase rotation to the columns of $\mathbf{B}$ without affecting the SINR.

### 3.1. Imperfect CSI

If the transmitter's knowledge of CSI is imperfect, then the precoder designed based on the above formulation in the assumption of perfect CSI may fail to achieve the required SINR. Here, we consider the design of precoders which will meet the SINR requirements of all users even when the CSI at the transmitter is imperfect.

#### 3.1.1. Channel Error Model

In the present context, we consider the situation where the transmitter CSI $\widehat{\mathbf{h}}_k$ is related to the true channel $\mathbf{h}_k$ as

$$\mathbf{h}_k = \widehat{\mathbf{h}}_k + \mathbf{e}_k. \qquad (7)$$

In one model, $\widehat{\mathbf{h}}_k$ is an imperfect estimate of the true channel $\mathbf{h}_k$ and $\mathbf{e}_k$ is a vector of i.i.d complex Gaussian random variables. This model is applicable when the user estimates its own channel and feeds it back to the transmitter through an ideal feed back link with no quantization, or if there is a delay between the estimation and the actual channel in fast varying environments.

In another model, $\widehat{\mathbf{h}}_k$ is a quantized version of the actual channel $\mathbf{h}_k$, and $\mathbf{e}_k$ represents the quantization error. This model is applicable when the user which knows the perfect channel $\mathbf{h}_k$ quantizes it and feeds back to the transmitter through a digital feedback link. Equation (8) can be used to model the uncertainty region in this case also.

In the robust precoder design, we consider the set

$$\mathcal{Z}_k = \{\mathbf{h}_k | \mathbf{h}_k = \widehat{\mathbf{h}}_k + \mathbf{e}_k, \|\mathbf{e}_k\| \leq \delta_k\}. \qquad (8)$$

The *uncertainty region* $\mathcal{Z}_k$ is the set of all channel vectors which lie in a sphere of radius $\delta_k$ around the estimated channel vector $\widehat{\mathbf{h}}_k$. This characterization of the uncertainty region is also related to the outage probability [13].

*3.1.2. Robust Precoder Design*

When the CSI at the transmitter is known to be imperfect, a robust precoder is designed to meet the target SINR of all the users. When the imperfections in the CSI are of the types described above, the robustness requirement of the precoder can be represented, in terms of the SOCP formulation as

$$\begin{aligned}
\min_{\mathbf{B}} \quad & \tau \\
\text{subject to} \quad & \|\mathbf{b}\| - \tau \leq 0, \\
& \max_{\mathbf{h}_k \in \mathcal{Z}_k} \left( \|[\mathbf{h}_k \mathbf{B} \ \sigma_k]\| - \mathbf{a}_k \mathbf{h}_k \right) \leq 0, \\
& 1 \leq k \leq N_u.
\end{aligned} \quad (9)$$

This problem is akin to the Robust Optimization (RO) [18], which is one of the methodologies for solving optimization problems under parameter uncertainties.

The general problem of optimization under parameter uncertainties has the following form:

$$\begin{aligned}
\min \quad & f_0(\boldsymbol{\zeta}) \\
\text{subject to} \quad & f_i(\boldsymbol{\zeta}, \mathbf{d}) \leq 0, \quad \forall \mathbf{d} \in \mathcal{R}, 1 \leq i \leq m,
\end{aligned} \quad (10)$$

where $\boldsymbol{\zeta} \in \mathbb{R}^n$ is the vector of decision variables, $\mathbf{d} \in \mathbb{R}^k$ are data vectors, $f_i$ are the constraints, and $\mathcal{R}$ is the *uncertainty set*. Computationally tractable approaches to the problem (10) for different classes of constraint functions $f$ and uncertainty regions $\mathcal{R}$ are reported in the literature [19]. For the case of robust precoder design (9), it is possible to use directly the results reported in [20] after converting the Second Order Cone (SOC) constraints to equivalent semi-definite constraints [21]. This approach is adopted in [12]. But the resulting robust counterpart is a Semi Definite Program (SDP) which is of higher complexity and computationally demanding than the SOCP formulation of the perfect CSI precoding problem.

Recent results have shown that its possible to have robust counterparts which preserve the structure of the nominal problem [22]. For the precoder design, this means that the robust design is a SOCP problem, which is a significant improvement over the SDP formulation. In this context, consider the data perturbation model

$$\mathbf{d} = \mathbf{d}^0 + \sum_{j \in N} \boldsymbol{\Delta} \mathbf{d}^j z_j, \quad (11)$$

where $\mathbf{d}^0$ is the nominal data value, $\boldsymbol{\Delta} \mathbf{d}^j$ are the directions of data perturbations, and $\{z_j, 1 \leq j \leq N\}$ are the zero mean i.i.d random variables. Robust optimization aims at finding a robust optimal $\boldsymbol{\zeta}$ which will meet the following constraint:

$$\max_{\mathbf{d} \in \mathcal{R}_\Omega} f(\boldsymbol{\zeta}, \mathbf{d}) \leq 0, \quad (12)$$

where

$$\mathcal{R}_\Omega = \left\{ \mathbf{d}_0 + \sum_{j \in N} \boldsymbol{\Delta} \mathbf{d}^j u_j \,\Big|\, \|\mathbf{u}\| \leq \Omega \right\}. \quad (13)$$

The following linearized version of the constraint (12) is considered in [22]:

$$\max_{(\mathbf{v},\mathbf{w}) \in \mathcal{V}_\Omega} f(\boldsymbol{\zeta}, \mathbf{d}^0) + \sum_{j \in N} \{ f(\boldsymbol{\zeta}, \boldsymbol{\Delta}\mathbf{d}) v_j + f(\boldsymbol{\zeta}, -\boldsymbol{\Delta}\mathbf{d}) w_j \} \leq 0, \quad (14)$$

where $\mathcal{V}_\Omega = \left\{ (\mathbf{v}, \mathbf{w}) \in \mathbb{R}_+^{2|N|} \,\Big|\, \|\mathbf{v} + \mathbf{w}\| \leq \Omega \right\}$.

It is shown in [22] that, for an SOC constraint, $\boldsymbol{\zeta}$ is feasible in (12) if $\boldsymbol{\zeta}$ is feasible in (14). We state the following theorem for the specific case of SOCP constraints.

**Theorem 1** (*Bertsimas-Sim [22]*)
*a) Constraint (14) is equivalent to*

$$f(\boldsymbol{\zeta}, \mathbf{d}^0) + \Omega \|\mathbf{s}\| \leq 0, \quad (15)$$

*where $s_j = \max\left\{ f(\boldsymbol{\zeta}, \boldsymbol{\Delta}\mathbf{d}^j), f(\boldsymbol{\zeta}, -\boldsymbol{\Delta}\mathbf{d}^j) \right\}$.*
*b) Equation (15) can be written as $\exists (y, \mathbf{t}) \in \mathbb{R}^{|N|+1}$*

$$\begin{aligned}
f(\boldsymbol{\zeta}, \mathbf{d}^0) &\leq -\Omega y & (16a) \\
f(\boldsymbol{\zeta}, \boldsymbol{\Delta}\mathbf{d}) &\leq t_j \quad \forall j \in N & (16b) \\
f(\boldsymbol{\zeta}, -\boldsymbol{\Delta}\mathbf{d}) &\leq t_j \quad \forall j \in N & (16c) \\
\|\mathbf{t}\| &\leq y. & (16d)
\end{aligned}$$

The following transformation [12] will enable us to write the precoder design problem in real variables:

$$\begin{aligned}
\overline{\mathbf{B}} &= \begin{bmatrix} \text{Re}(\mathbf{B}) & \text{Im}(\mathbf{B}) \\ -\text{Im}(\mathbf{B}) & \text{Re}(\mathbf{B}) \end{bmatrix}, & (17) \\
\overline{\mathbf{h}}_k &= [\text{Re}(\mathbf{h}_k) \quad \text{Im}(\mathbf{h}_k)], & (18) \\
\overline{\mathbf{b}}_k &= [\text{Re}(\mathbf{b}_k) \quad -\text{Im}(\mathbf{b}_k)], & (19) \\
\overline{\mathbf{e}}_k &= [\text{Re}(\mathbf{e}_k) \quad \text{Im}(\mathbf{e}_k)]. & (20)
\end{aligned}$$

The precoder design problem can be formulated in terms of real variables by replacing the complex vectors and matrices in (9) by the corresponding real vectors and matrices obtained by the transformations given above.

The data perturbation model (11) for the second order cone constraint of the precoder design problem takes the form

$$\mathbf{d}_k = \mathbf{d}_k^0 + \sum_{N_u} \boldsymbol{\Delta}\mathbf{d}_k^j \bar{e}_{k,j}, 1 \leq k \leq N_u, 1 \leq j \leq 2N_t. \quad (21)$$

where $\mathbf{d}_k = [\overline{\mathbf{h}}_k \ \overline{\mathbf{h}}_k]^T$, $\mathbf{d}^0 = [\overline{\mathbf{h}}_k \ \overline{\mathbf{h}}_k]^T$, $\boldsymbol{\Delta}\mathbf{d}_k^j = [\mathbf{i}_j \ \mathbf{i}_j]^T$, and $\mathbf{i}_j$ is the $j$th row of $2N_t \times 2N_t$ identity matrix. $\mathbf{d}$ is the vector of all data in the problem and has the structure given above as $\overline{\mathbf{h}}_k$ appears twice in the constraint in (9). $\boldsymbol{\Delta}\mathbf{d}_k^j$ indicates how the error in the $j$th component of $\mathbf{h}_k$ affects $\mathbf{d}$. Based on this data perturbation model, it is obvious that the channel uncertainty region $\mathcal{Z}_k$ of the robust precoder design problem of (9) corresponds to the uncertainty region in (13), with $\Omega = \delta_k$.

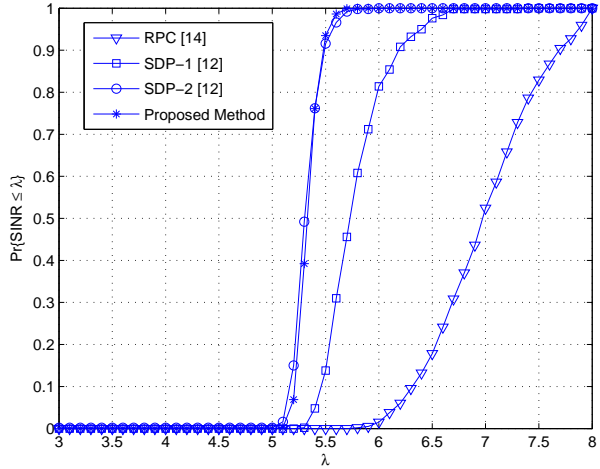

**Fig. 1**. CDF of achieved SINR at the downlink users. Minimum required SINR $\gamma_1 = \gamma_2 = \gamma_3 = 5$ dB. $N_t = N_u = 3$, uncertainty size, $\delta_1 = \delta_2 = \delta_3 = 0.015$.

Using the data perturbation model in (21) and applying Theorem-1 to (9), we obtain the following SOCP formulation of the proposed robust precoder design:

$$\min_{\mathbf{B}} \quad \tau \tag{22a}$$

$$\text{subject to} \quad \left\|\overline{\mathbf{b}}_k\right\| \leq \tau \tag{22b}$$

$$\left\|\begin{bmatrix}\overline{\mathbf{h}}_k^T\overline{\mathbf{B}} & \sigma_k\end{bmatrix}\right\| - a_k\overline{\mathbf{h}}_k^T\overline{\mathbf{b}}_k \leq -\delta_k y_k \tag{22c}$$

$$\left\|\begin{bmatrix}\overline{\mathbf{b}^i}^T & \sigma_k\end{bmatrix}\right\| - a_k\overline{\mathbf{B}}_{i,k} \leq t_{k,i} \tag{22d}$$

$$\left\|\begin{bmatrix}\overline{\mathbf{b}^i}^T & \sigma_k\end{bmatrix}\right\| + a_k\overline{\mathbf{B}}_{i,k} \leq t_{k,i} \tag{22e}$$

$$\|\mathbf{t}_k\| \leq y_k \tag{22f}$$

$$1 \leq k \leq N_u, \quad 1 \leq i \leq 2N_t. \tag{22g}$$

where $\mathbf{b}^i$ is the $i$th row of $\mathbf{B}$. In this formulation of the robust precoder design, the constraints are of the same type of the nominal problem (11). Hence, the computational complexity is of the same order as the nominal problem.

The robustness constraint in (14) is a relaxation of the constraint in (12). By selecting appropriate value of $\kappa, 0 \leq \kappa \leq 1$, and replacing $\delta_k$ in (22) by $\kappa\delta_k$, it is possible to get a robust precoder which transmits less power while achieving the required SINR constraints. Through extensive simulations, it was found that $\kappa = 1/4$ provides good balance between the achieved SINR and the transmit power.

## 4. SIMULATION RESULTS

In this section, we present the performance of the proposed robust precoder design (22) through simulations. We compare this performance with some other robust designs available in the literature. The components of the estimated chan-nel vectors $\widehat{\mathbf{h}}_k$, $1 \leq k \leq N_u$ are i.i.d zero mean unit variance proper complex Gaussian random variables. The noise samples $n_k$, $1 \leq k \leq N_u$ are i.i.d proper complex Gaussian random variables with zero mean and unit variance. We compare the performance of the proposed design with the robust SOCP design (denoted here by SDP-1) and the unstructured SDP design (denoted by SDP-2) in [12], and the robust power control (denoted by RPC) in [14].

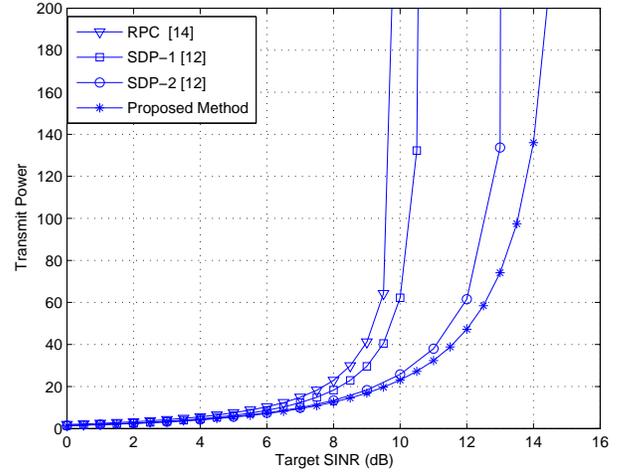

**Fig. 2**. Transmit power versus SINR requirement of the users. Uncertainty size $\delta = \delta_1 = \delta_2 = \delta_3 = 0.02$, $N_t = N_u = 3$.

First, we compare the CDF of the achieved SINR of SDP-1, SDP-2, and RPC with the CDF of SINR of the proposed robust design. Figure 1 shows the CDF for various methods. In this experiment, we consider a system with a base station having $N_t = 3$ transmit antennas and $N_u = 3$ single antenna receivers. The uncertainty size of CSI at the transmitter is assumed to be same for all users and is $\delta = 0.015$. The target SINR for all users is $\gamma = 5$ dB. In case of SDP-1 and RPC, it is evident that, most of the time, the users get SINR much higher than the target SINR. This implies that these algorithms result in much higher transmit power than required. The SDP-2 and the proposed design have almost same CDF, and is very near to the required SINR. That is, performance wise, the proposed design achieves almost the same performance as SDP-2 in [12] but with reduced complexity.

Figure 2 shows the transmit power $\text{Tr}\{\mathbf{B}^H\mathbf{B}\}$ for various robust designs in order to achieve different target SINRs. This experiment also has the same setting as above, except for the target SINR which is varied from 0 dB to 10 dB. The SDP-1 and RPC methods transmits more power compared to SDP-2 and the proposed method. This higher transmit power results in the higher SINRs at the users.

Figure 3 shows the transmit power for the different robust designs for different values of the size of channel uncertainty. The SINR requirement for all users is 5 dB. The SDP-1 and RPC methods end up in higher transmit power compared to SDP-2 and the proposed method. This higher transmit power results in the higher SINRs at the users. Also the range of $\delta$ for which the proposed method is feasible is larger than other methods.

**Table 1**. Comparison of Run-Time in Seconds for different precoding methods

| Method | $N_u,N_t$=3 | $N_u,N_t$=4 | $N_u,N_t$=5 | $N_u,N_t$=6 |
|---|---|---|---|---|
| Proposed | 0.10 | 0.2 | 0.44 | 0.6 |
| SDP-1 [12] | 0.2 | 0.3 | 0.6 | 1 |
| SDP-2 [12] | 4.5 | 16 | 61 | 121 |

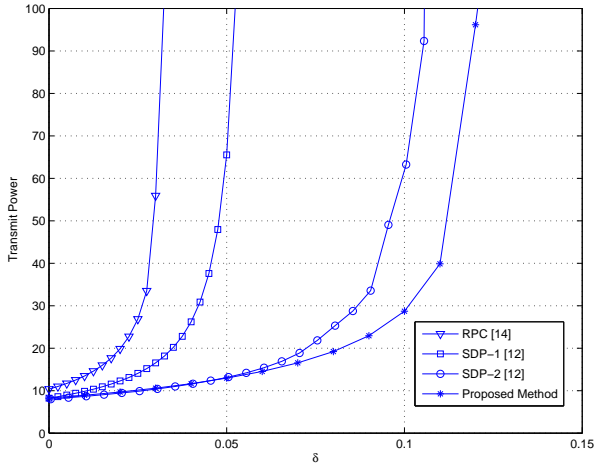

**Fig. 3**. Transmit power versus channel uncertainty size, $\delta$. $N_t = N_u = 3$, SINR requirement of the users $\gamma_1 = \gamma_2 = \gamma_3 = 5$ dB.

Table-1 shows the comparison of computation time in seconds required for solving the robust precoder using different methods on a 2.66 GHz machine using the solver SeDuMi [17]. Computation time for SDP-2 is the highest. Computation time for SDP-1 and the proposed method are comparable. The proposed method is able to achieve the performance comparable to SDP-2 at the computational cost of SDP-1.

In summary, the proposed robust design achieves better performance than the other methods compared while being computationally less intensive.

## 5. CONCLUSION

We proposed a design of robust precoder with SINR constraints for multiuser MISO downlink with imperfect CSI at the transmitter. We showed that the robust precoder design problem can be formulated as a SOCP. The SOCP formulation has the advantage that the computational complexity of the robust design is of the same order as that of the design with perfect CSI. A comparison with other robust precoder designs reported in the literature showed that the proposed robust design performs better while being computationally less complex.